\documentclass[review=false]{acmart}

\usepackage{multirow}
\usepackage[ruled,vlined]{algorithm2e}
\AtBeginDocument{%
  }

\copyrightyear{2023}
\acmYear{2023}
\setcopyright{rightsretained}
\acmConference[SIGITE '23]{The 23rd Annual Conference on Information Technology Education}{October 11--14, 2023}{Marietta, GA, USA}
\acmBooktitle{The 23rd Annual Conference on Information Technology Education (SIGITE '23), October 11--14, 2023, Marietta, GA, USA}
\acmDOI{10.1145/3585059.3611431}
\acmISBN{979-8-4007-0130-6/23/10}

\begin{document}

\title[ChatGPT for Teaching and Learning in Data Science]{ChatGPT for Teaching and Learning:\\An Experience from Data Science Education}

\author{Yong Zheng}
\affiliation{%
  \institution{Illinois Institute of Technology}
  \city{Chicago}
  \state{Illinois}
  \country{USA}
  \postcode{60616}
}
\email{yzheng66@iit.edu}

\begin{abstract}
ChatGPT, an implementation and application of large language models, has gained significant popularity since its initial release. Researchers have been exploring ways to harness the practical benefits of ChatGPT in real-world scenarios. Educational researchers have investigated its potential in various subjects, e.g., programming, mathematics, finance, clinical decision support, etc. However, there has been limited attention given to its application in data science education. This paper aims to bridge that gap by utilizing ChatGPT in a data science course, gathering perspectives from students, and presenting our experiences and feedback on using ChatGPT for teaching and learning in data science education. The findings not only distinguish data science education from other disciplines but also uncover new opportunities and challenges associated with incorporating ChatGPT into the data science curriculum.

\end{abstract}


\maketitle

\section{Introduction}
\noindent
The large language model (LLM) is a type of artificial intelligence model designed to understand and generate human-like text based on the patterns it has learned from extensive training on vast amounts of data. The pre-trained LLMs can be obtained by training from large texts data, e.g., books, articles, Webpages, etc. They are able to capture the statistical patterns and structures of human language, allowing the model to learn grammar, syntax, and semantic relationships. Later, LLMs undergo fine-tuning tailored to particular tasks or domains, e.g., the task of machine translation~\cite{brants2007large}, text classifications~\cite{howard2018universal}, text generation~\cite{li2021pretrained}, etc.

The Generative Pre-trained Transformer (GPT) is among the widely recognized LLMs, evolving from GPT-1 to the recent GPT-4. OpenAI has developed ChatGPT as an implementation of GPT customized for conversational interactions. ChatGPT, based on the powerful GPT-3 model, benefits from its impressive scale of 175 billion parameters, which provides the foundation and capabilities necessary for dynamic and interactive conversations with users.

ChatGPT has gained significant popularity due to several factors, e.g., the capability of content generations, friendly user interfaces for question-answering, the public availability, the potential for practical usage and applications, etc. Researchers in the education domain also made several attempts to seek opportunities of using ChatGPT for teaching and learning. Various functions by ChatGPT have been identified to assist and improve education~\cite{lo2023impact,rospigliosi2023artificial,biswas2023role,firat2023chatgpt}, such as personalized learning, concept understanding, coding generation and explanations, educational assessment, etc.

Based on our literature reviews, researchers have extensively explored the potential of ChatGPT in educational teaching and learning for different subjects, like computer programming~\cite{chen2023gptutor,biswas2023role,surameery2023use}, mathematics~\cite{wardat2023chatgpt,sanchez2023chatgpt}, economics and finance~\cite{geerling2023economics,geerling2023chatgpt}, medical education~\cite{khan2023chatgpt,lee2023rise,arif2023future}, and so on. They have also revealed that the effectiveness and usage of ChatGPT may vary from subjects to subjects~\cite{lo2023impact}. Meantime, we realize that there is a noticeable gap in the application of ChatGPT within data analytics and data science education. The current literature lacks substantial attention to how ChatGPT can be effectively utilized in this specific domain. 

This paper seeks to address this gap by incorporating ChatGPT into a data science course and collecting valuable insights from students, as well as sharing feedback from the instructor. The contributions in this paper can be summarized as follows. 
As far as we know, this is the first attempt to reveal the opportunities and challenges of using ChatGPT for data science education through real-world practice on data science curriculum and user studies from graduate students. Moreover, this study helps distinguish data science education from other subjects, and identify new opportunities or challenges of using ChatGPT for data science education. Finally, the findings from this research endeavor and the user studies from graduate students have the potential to enhance the educational landscape by exploring ChatGPT in data science education and informing future educational practices in this field.


\section{Related Work}

\subsection{ChatGPT for Education: Opportunities}
Researchers have identified multiple opportunities of using ChatGPT for education. First of all, the benefits for learners include but are not limited to:

\begin{itemize}
	\item \textit{Personalized Learning}~\cite{rahman2023chatgpt, rospigliosi2023artificial,biswas2023role,kasneci2023chatgpt}. With ChatGPT, students can ask questions in their own words and receive responses tailored to their specific chat sessions~\cite{rospigliosi2023artificial}.
	\item \textit{Tutoring and Assistance}~\cite{biswas2023role,lo2023impact,kasneci2023chatgpt}. ChatGPT can help summarize information, clarify and understand concepts, generate questions for practice through the question-answering interactions~\cite{lo2023impact}. Particularly, ChatGPT can provide enriched support in coding or programming, such as coding generations, error checking/debugging, coding explanations, etc.~\cite{rahman2023chatgpt}.
	\item \textit{Reading/Writing Assistance and Improvements}~\cite{kasneci2023chatgpt,biswas2023role}. In terms of the reading, ChatGPT is able to explain a term or concept in different ways, especially for junior students~\cite{kasneci2023chatgpt}. ChatGPT is much more powerful for assisting writing, not only for children or junior students, but also for essay and paper writing by graduate students~\cite{kasneci2023chatgpt,biswas2023role}, e.g., paper proofreading, grammar check, sentence rephrasing, paper editing, etc.
	\item \textit{Support for Group Learning and Collaborations}~\cite{rudolph2023chatgpt,kasneci2023chatgpt}. The latest version of ChatGPT allows users to share conversations with others, which results in convenience in facilitating student collaborations, group learning and discussions.
	\item \textit{Critical Thinking and Problem-Solving}~\cite{kasneci2023chatgpt,halaweh2023chatgpt}. There are several debates or conflicting opinions on fostering critical thinking and problem-solving skills. On one hand, students can read the answers from ChatGPT, compare them with the answers in their own mind, thus they can improve critical thinking and problem-solving skills. On  the other hand, students may directly acquire answers from ChatGPT without learning from these answers.
\end{itemize}

Moreover, there are also benefits for educators, including generating or suggesting course materials~\cite{lo2023impact,kasneci2023chatgpt}, lesson planning (e.g., generating syllabus)~\cite{rahman2023chatgpt,kasneci2023chatgpt}, language translation for teaching materials~\cite{lo2023impact}, generating assessment items (e.g., quizzes, exercises, etc.)~\cite{lo2023impact}, evaluating student performance (e.g., essay grading)~\cite{lo2023impact,rahman2023chatgpt,kasneci2023chatgpt}, support for virtual office hours~\cite{biswas2023role} and QA sessions~\cite{rahman2023chatgpt}, assisting research writing~\cite{kasneci2023chatgpt}, and so forth.

\subsection{Variations in Different Subjects}
Lo et al. revealed that the performance or effectiveness of ChatGPT may vary across subject domains through their literature review and summaries~\cite{lo2023impact}. For example, they found that ChatGPT could deliver outstanding performance for economics and finance, but the overall performance may stay in the range between "outstanding" and "satisfactory" for computer programming tasks. By contrast, the performance could be barely satisfactory for applications in law and medical education. When it comes to psychology and mathematics education, there would be more improvements needed for ChatGPT to deliver satisfied answers or outcomes. Consequently, we believe that the application of ChatGPT in one particular subject or domain cannot be seamlessly transferred or replicated to another. This has motivated us to undertake our own investigation, seeking to uncover the unique opportunities and challenges associated with employing ChatGPT in the context of data science education.

\section{Research Problems and Study Design}
In this section, we first introduce data science education and compare it with other subjects, which leads to the statement of research problems in this paper. We also introduce the design of our studies, as well as evaluation strategies.

\subsection{Data Science vs. Other Subjects}
Within our information technology graduate program, the data science curriculum comprises three essential mandatory courses -- a database and data management course, an introductory data analytics course, and a data mining and machine learning course. It is important to note that the database course is not specifically discussed in this context, as it is a general course that encompasses various specializations such as software engineering, web design, and cybersecurity, rather than for 
data science education only. 


Hence, our specific focus lies on the data analytics and data science courses. The data analytics class primarily imparts knowledge and skills related to statistical analysis and linear regression models, while the data science classes delve further into the realm of data preprocessing, practical techniques for data mining, and machine learning. Despite their distinct areas of emphasis, both courses share similar teaching styles and learning objectives. More specifically, they introduce fundamental concepts and knowledge in data analytics and data science, and also deliver practical data science programming skills, such as Python or R for data science/analytics. Students are expected to be able to apply their acquired knowledge and skills to solve real-world problems. As a culminating assessment, they are tasked with completing a final project wherein they must locate a real-world dataset, propose problem statements to be addressed, and employ their practical skills to devise multiple solutions and evaluate their efficacy.

To distinguish data science from other subjects or courses, we wish to emphasize two crucial aspects: the key skills that define data science and the role of programming within this field. 

\begin{itemize}
	\item Firstly, it is important to highlight the role of programming skills in data science curricula. Students must possess programming proficiency, particularly in languages like Python, to conduct data analysis and problem-solving tasks. These tasks encompass various activities such as data preprocessing, data visualization, hypothesis testing, building and evaluating machine learning models, among others. However, it is noteworthy that, in most instances, students primarily focus on preparing inputs, utilizing built-in functions available in Python libraries (e.g., scikit-learn), and tuning hyperparameters within these functions. The programming skills required in data science curricula are typically less complex compared to those in other subjects such as software engineering or web design. For example, students may not even need to use \textit{if-else} or \textit{loops} in their practice.
	\item Furthermore, the essential skills in data science extend beyond mere programming proficiency. The capabilities revolve around effective problem-solving. When presented with real-world datasets, students are required to make informed decisions regarding the data types involved and employ appropriate preprocessing techniques to cleanse and prepare the data for analysis. Similarly, when confronted with practical problems, students must identify the most suitable approach to transform the problem into a data science task. Decisions regarding which solutions to utilize and how to accurately evaluate these solutions also fall within the realm of their responsibilities. In essence, data science education emphasizes the development of problem-solving skills, enabling students to tackle complex challenges and derive meaningful insights from data.
\end{itemize}

\subsection{The Design of Our Studies}
Therefore, the major research problem in this paper is to exploit how well ChatGPT can help solve these issues (e.g., decision making, problem solving, data preprocessing, etc.) in data science, and how students can benefit from ChatGPT to assist learning in ChatGPT. 

In our studies, we design and deploy a practice of using ChatGPT based on GPT-3.5. Students need to use ChatGPT to enter their own prompts to acquire the answers. There are several practices classified into 7 scenarios in this study. We had a specific goal in each scenario, e.g., examining the capability of coding generations. Students need to complete them one by one, and also complete a questionnaire. The questionnaire is composed of 10 questions which correspond to their previous practice. The practice and corresponding questions in our study can be described as follows.

\subsubsection{Scenario 1: Using ChatGPT for general programming}\leavevmode\\
\textit{Practice: }ask ChatGPT to give you the coding of popular algorithms (e.g., bubble sort, vector space model, etc.) in any selected programming language (e.g., Python, C++, Java, etc.)

	\textit{Goal: }it is used to exploit the capability of coding generations in general (i.e., not specific to coding for data science). Later, students can try coding practice on data science tasks.
	
	\textit{Q1. How much do you agree that ChatGPT can produce high-quality programming coding with no or less human effort? }

\subsubsection{Scenario 2: Concept and knowledge learning in data science}\leavevmode\\
	\textit{Practice: }ask ChatGPT to help you understand some terms or concepts, such as new terms like "active learning" which was not introduced in the class, and old terms like "imbalance issue in the classification task" which was covered by the class.  Students can even ask ChatGPT to given answers in their native language\footnote{Note: more than half of the students in the class were international students}.
	
	\textit{Goal: }examine whether ChatGPT can assist concept/knowledge understanding
	
	\textit{Q2. How much do you agree that ChatGPT can help you better understand new concepts or knowledge in data science?}
	
	\textit{Q3. How much do you agree that ChatGPT can help you better understand or clarify old concepts or knowledge in data science?}
	
\subsubsection{Scenario 3: Data analysis without human effort}\leavevmode\\
	\textit{Practice: ask ChatGPT to build a machine learning model (e.g., decision tree model) on a specific data set (e.g., Iris data or any data set on Kaggle.com)}
	
	\textit{Goal: }examine whether ChatGPT can complete the data analytics or data science tasks without human effort.
	
	\textit{Q4. How much do you agree that ChatGPT can help analyze data sets with no or less human effort? (by considering whether it can automatically give correct preprocessing, or handle special issues, such as imbalance, automatically)}
	
\subsubsection{Scenario 4: Critical thinking and problem solving}\leavevmode\\
	\textit{Practice: }ask ChatGPT to help solve a specific problem, e.g., how to deal with imbalance issue, how to alleviate overfitting in decision trees, how to find the optimal K value for K-Means clustering. 
	
	\textit{Goal: }examine whether ChatGPT can offer problem-solving solutions with appropriate prompts.
	
	\textit{Q5. How much do you agree that ChatGPT can help you foster critical thinking and problem solving skills?}

\subsubsection{Scenario 5: Coding for data science}\leavevmode\\
	\textit{Practice: }ask ChatGPT to generate Python coding for data science tasks, e.g., build a decision tree model on Iris data set, build decision trees with solutions to alleviate overfitting, etc. Students were also asked to work on more practice, e.g., ask ChatGPT to explain the coding, let ChatGPT to explain the hyperparameters in a specific Python function.
	
	\textit{Goal: }examine the capability of coding generations for data science tasks, as well as seeking more assistance in coding explanations or API explanations. Students can give more prompts, e.g., they want to use a specific strategy to alleviate overfitting. Note that this practice is different from scenario 3.2.3 where we try to examine whether ChatGPT can generate correct coding without human effort.
	
	\textit{Q6. How much do you agree that ChatGPT can help you use or generate Python coding for data science, with appropriate human efforts or prompts.}
	
	\textit{Q7. How much do you agree that ChatGPT can help you understand or explain Python coding for data science}
	
\subsubsection{Scenario 6: Learning material suggestions}\leavevmode\\
	\textit{Practice: }ask ChatGPT to suggest some learning materials (e.g., textbooks, YouTube videos, social media accounts, etc.). Moreover, students were asked to examine whether ChatGPT can suggest appropriate Python libraries or tools for data science, e.g., libraries for alleviating imbalance issues, tools for face detection or emotion recognition, and so forth.
	
	\textit{Goal: }examine the appropriateness of suggested materials.
	
	\textit{Q8. How much do you agree that ChatGPT can suggest appropriate learning materials to you (by also considering whether these materials are real ones)}
	
	\textit{Q9. How much do you agree that ChatGPT can suggest appropriate or useful Python libraries for data science)}
	
\subsubsection{Scenario 7: Recommendation of skill sets and career path}\leavevmode\\
	\textit{Practice: }ask ChatGPT to suggest some skill sets or career path. For example, "I have learned R programming for linear regressions, what knowledge and skills I need to learn in order to find a data scientist position?"
	
	\textit{Goal: }examine the advice by ChatGPT for skill sets/career path.
	
	\textit{Q10. How much do you agree that ChatGPT can suggest appropriate skill sets or career path to you.}

\leavevmode\\	
For all the questions above, students need to give answer in 1 (strongly disagree) to 5 (strongly agree) scale in the questionnaire. 

Moreover, the instructor (i.e., the author) also completed these practices, so that we can deliver the answers from the perspectives of both students and the instructor. In addition to the benefits for learners in data science education, the instructor also tried different ways to seek help from ChatGPT for teaching. The initial findings will also be discussed in Section 4.

\section{Results and Findings}
There were 28 students who completed the practice and questionnaire. We discuss the benefits for learners from the perspectives of both students and instructors in Section 4.1. The attempts to seek benefits for education were presented in Section 4.2.

\subsection{Benefits for Learners}
\begin{figure*}
  \includegraphics[width=\linewidth]{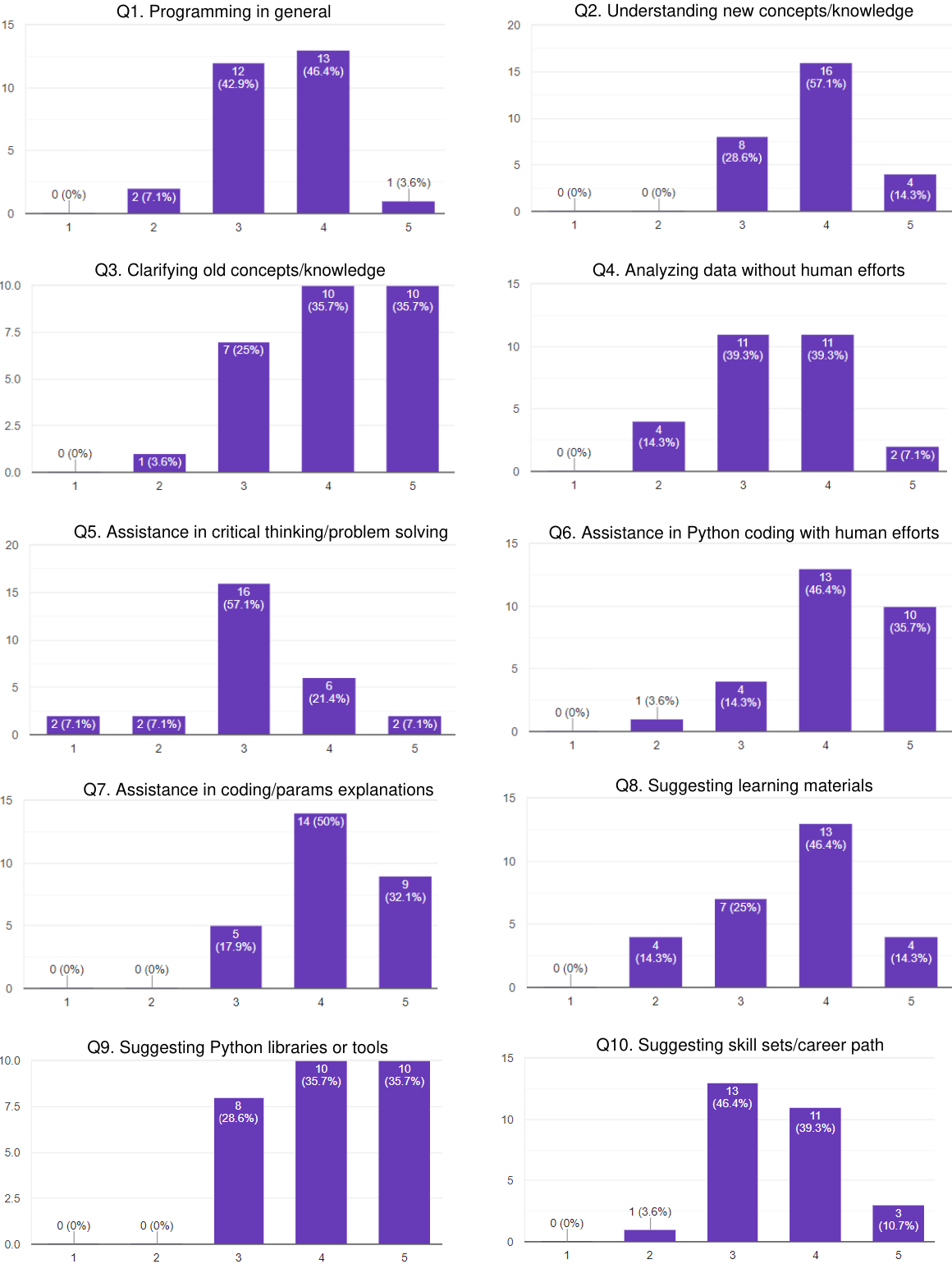}
  \caption{Perspectives from Students}\label{fig:answers}
\end{figure*}

Students' answers in the questionnaire can be visualized by Figure~\ref{fig:answers}. The x-axis represents the scale 1 (strongly disagree) to 5 (strongly agree). The y-axis tells the number of student votes. In the bars, we present both the number of votes and the ratio in percentage values. We summarize the answers by students and indicate the perspective of the instructor as follows.\\

\noindent
\textit{Q1. Programming in general}
\begin{itemize}
	\item Perspective of students: only half of the students gave satisfactory results, and they believe ChatGPT can help produce programming codes. However, there were 13 students who gave a "neural" answer and 2 students disagreed with ChatGPT's performance in programming.
	\item Perspective of the instructor: the answer may vary from the specific practice. If a well-known algorithm is given to ChatGPT, such as bubble sort, ChatGPT can produce perfect answers. Once the algorithm is more complex in the query, ChatGPT may not work very well.
\end{itemize}

\noindent
\textit{Q2. Understanding new concepts/knowledge}
\begin{itemize}
	\item Perspective of students: most students believe that ChatGPT is helpful in understanding new concepts or knowledge.
	\item Perspective of the instructor: When it comes to a new concept or term, students actually do not know the true answer. Therefore, students or users can only give their subjective opinions on the textual answers produced by ChatGPT.
\end{itemize}

\noindent
\textit{Q3. Clarifying old concepts/knowledge}
\begin{itemize}
	\item Perspective of students: most students concluded its usefulness in clarifying old concepts or knowledge, though there was one student who gave a "disagree" answer.
	\item Perspective of the instructor: The prompt given to ChatGPT is the key to obtaining useful answers. If one prompt does not work well, it is suggested to try another one. ChatGPT can give appropriate answers customized to students, as long as they can try with multiple prompts. If a student does not understand the answers in this way, it is possible that they can understand another answer by trying other prompts.
\end{itemize}

\begin{figure}[htb]
\centering
  \includegraphics[scale=0.55]{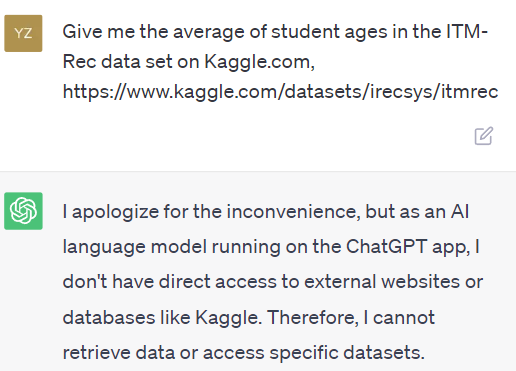}
  \caption{Example of Using External Data}\label{fig:itmdata}
\end{figure}

\noindent
\textit{Q4. Analyzing data without human efforts}
\begin{itemize}
	\item Perspective of students: 46.4\% of students considered ChatGPT as being helpful in analyzing data. It is probably that they tried the Iris data set only, without trying other data. 
	\item Perspective of the instructor: Iris is a well-known small benchmark data set for data science. Due to its small size, Iris data set is usually embedded in the software (e.g., Matlab) or data science library (e.g., scikit-learn). ChatGPT can produce appropriate data science coding (e.g., decision trees) on these built-in data sets. However, ChatGPT cannot take an external data set as inputs, or users cannot give a data file in a query, e.g., the example of using the ITM-Rec~\cite{zheng2023itmrec} data shown by Figure~\ref{fig:itmdata}. ChatGPT actually cannot analyze data set directly, if a data set is unknown. Beyond this point, ChatGPT cannot provide coding solutions for specific data science problems or challenges, such as alleviating imbalance issues, handling appropriate data preprocessing for a specific variable, and so forth. ChatGPT needs extra information (e.g., decisions or specific requests) from humans to correctly analyze data.
\end{itemize}

\noindent
\textit{Q5. Assistance in critical thinking/problem-solving}
\begin{itemize}
	\item Perspective of students: only 28.5\% of students believed ChatGPT was useful towards critical thinking or problem solving. 
	\item Perspective of the instructor: Given a specific problem, e.g., imbalance issues, ChatGPT is able to deliver solutions in general. However, it is another story when students tried to apply these solutions to a specific data set. Students may need to provide more information in the prompts in order to acquire the desired answers.
\end{itemize}

\noindent
\textit{Q6. Assistance in Python coding with human efforts}
\begin{itemize}
	\item Perspective of students: most students believed that ChatGPT can work well for Python coding in data science tasks, if appropriate human efforts (e.g., human decisions on evaluation strategies or solutions for imbalance issues) were given in the prompts.
	\item Perspective of the instructor: The appropriate coding for data science relies on human efforts. As indicated previously, ChatGPT may not figure out appropriate data preprocessing or solutions to handle specific issues or challenges, since these issues may be associated with a specific data set. Once humans or users are able to make decisions for problem-solving and give specific requests in the prompts, ChatGPT can produce better coding for data science tasks. 
\end{itemize}

\begin{figure}[htbp]
\centering
\begin{minipage}[t]{0.48\textwidth}
\centering
\includegraphics[scale=0.33]{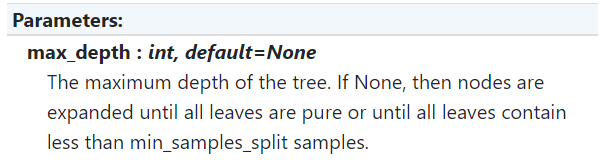}
\caption{Explaining `max\_depth' by API}\label{fig:api}
\end{minipage}
\begin{minipage}[t]{0.48\textwidth}
\centering
\includegraphics[scale=0.33]{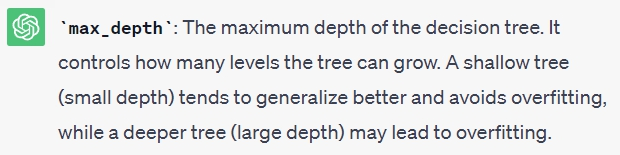}
\caption{Explaining `max\_depth' by ChatGPT}\label{fig:gpt}
\end{minipage}
\end{figure}

\noindent
\textit{Q7. Assistance in coding/params explanations}
\begin{itemize}
	\item Perspective of students: More than 80\% of students concluded its usefulness.
	\item Perspective of the instructor: We believe that ChatGPT delivers outstanding performance in coding or parameter explanations. Take hyperparameters for one example, the Python library provides official API to explain the hyperparameters in each Python function. However, students still had difficulties in understanding these parameters from the API, since API may only introduce the meaning, the scale or different options of a hyperparameter. By contrast, ChatGPT also tells the effects of hyperparameters, e.g., using a smaller value in this parameter may result in less running time but lower accuracy. By this way, students are cleared about which hyperparameters to tune up and how to change them. 
	
	\hspace{10pt}Figure~\ref{fig:api} and Figure~\ref{fig:gpt} show the example of explaining the parameter `max\_depth' in the DecisionTreeClassifier in the scikit-learn library by using official API and ChatGPT, respectively. The explanations by ChatGPT in Figure~\ref{fig:gpt} can additionally tell the effects by using a small or large depth value in this parameter.
\end{itemize}

\noindent
\textit{Q8. Suggesting learning materials}
\begin{itemize}
	\item Perspective of students: Around 60\% of students believed that ChatGPT was useful to suggest learning materials.
	\item Perspective of the instructor: ChatGPT may produce some relevant or useful information. However, the validity of these materials may be still in doubt. For example, ChatGPT may suggest a textbook with the wrong authors or publishers. Google search may work better than ChatGPT, at least Google search can return relevant and true information.
\end{itemize}

\noindent
\textit{Q9. Suggesting Python libraries or tools}
\begin{itemize}
	\item Perspective of students: More than 70\% of students concluded its usefulness.
	\item Perspective of the instructor: ChatGPT works well for suggesting useful Python libraries or tools. It is necessary, since not all solutions in data science are included in the standard Python libraries, e.g., scikit-learn.
\end{itemize}

\noindent
\textit{Q10. Suggesting skill sets/career path}
\begin{itemize}
	\item Perspective of students: Only half of the students believed that ChatGPT worked well for these suggestions.
	\item Perspective of the instructor: The correct skill sets or career path are dependent with student profiles, e.g., their background, previous learning history, learning or career goals, as well as the latest requirements by recruiters. ChatGPT may still have challenges in producing excellent answers at this moment.
\end{itemize}

\subsection{Benefits for Educators}
We tried to use ChatGPT to help prepare syllabuses, suggest teaching materials, generate simulated data set and produce quizzes or exercises. It is surprising that ChatGPT can work well in generating data sets. If the requested data set is small, e.g., 100 rows, ChatGPT can directly produce the data and print them out. If the requested data is large, e.g., 10,000 rows, ChatGPT will generate a piece of Python coding to help produce the simulated data. For generating data sets and exercises, users need to give more information in the prompts in order to help achieve high-quality outputs. When it comes to assessment or evaluations, e.g., grading, ChatGPT may not work well in data science education, since they cannot grade questions related to problem solving, while there could be multiple available and correct answers, rather than unique answers.

\section{Conclusions and Future Work}
\noindent
In this paper, we performed user studies to let students engage in practice of using ChatGPT for data science. Based on the studies, we revealed the useful functions or opportunities (e.g., coding and hyperparameter explanations) of using ChatGPT to assist learners in data science education, while we also identified some challenges, e.g., assistance in critical thinking or problem solving, suggesting learning materials or skill sets. There are some limitations in this work, e.g., the pool of students in user studies is small. Also, the studies were performed on the data science course only. 

In our future work, we will extend the studies to more other data science curriculum and have more students engaged in the studies in order to deliver enriched findings and results. Moreover, we also plan to seek optimal ways to incorporate ChatGPT in data science education, e.g., alleviating the ethic issues by limiting the usage in specific learning scenarios, figuring out new ways to utilize ChatGPT for assessment in data science education, and so forth.

\bibliographystyle{ACM-Reference-Format}
\bibliography{sample-base}

\end{document}